\documentclass[11pt]{article}
\baselineskip
\pdfoutput=1
\usepackage{amsmath}
\usepackage{amssymb}
\usepackage{graphicx}
\usepackage{color}
\usepackage{cite}
\usepackage{collref}
\usepackage{tabls}
\usepackage{hyperref}
\usepackage{diagbox}
\newcommand{\SU}{\mathrm{SU}}
\newcommand{\DKL}{\tilde{D}_{\mbox{\tiny{KL}}}}
\newcommand{\dd}{{\rm{d}}}
\newcommand{\kB}{k_{\mbox{\tiny{B}}}}
\newcommand{\etain}{\eta_{\mbox{\tiny{in}}}}
\newcommand{\etafin}{\eta_{\mbox{\tiny{fin}}}}
\newcommand{\tin}{t_{\mbox{\tiny{in}}}}
\newcommand{\tfin}{t_{\mbox{\tiny{fin}}}}
\newcommand{\nab}{n_{\mbox{\tiny{ab}}}}
\newcommand{\nsb}{n_{\mbox{\tiny{sb}}}}
\newcommand{\nmeas}{N_{\mbox{\tiny{meas}}}}

\newcommand{\eq}[1]{\begin{equation}\label{#1}}
\newcommand{\en}{\end{equation}}
\newcommand{\eqar}[1]{\begin{eqnarray}\label{#1}}
\newcommand{\enar}{\end{eqnarray}}

\begin {document}

\begin{titlepage}

\begin{center}
{\Large\bf Stochastic normalizing flows as non-equilibrium transformations}
\end{center}
\vskip1.3cm
\centerline{Michele~Caselle$^{1,2}$,\footnote{\href{mailto:caselle@to.infn.it}{{\tt caselle@to.infn.it}}} Elia~Cellini$^{1,2}$,\footnote{\href{mailto:elia.cellini@unito.it}{{\tt elia.cellini@unito.it}}} Alessandro~Nada$^{1}$\footnote{\href{mailto:alessandro.nada@unito.it}{{\tt alessandro.nada@unito.it}}} and Marco~Panero$^{1,2}$\footnote{\href{mailto:marco.panero@unito.it}{{\tt marco.panero@unito.it}}}}
\vskip1.5cm
\centerline{\sl $^{1}$ Department of Physics, University of Turin}
\centerline{\sl $^{2}$ INFN, Turin}
\centerline{\sl Via Pietro Giuria 1, I-10125 Turin, Italy}
\vskip1.0cm

\setcounter{footnote}{0}

\begin{abstract}
\noindent
Normalizing flows are a class of deep generative models that provide a promising route to sample lattice field theories more efficiently than conventional Monte~Carlo simulations. In this work we show that the theoretical framework of stochastic normalizing flows, in which neural-network layers are combined with Monte~Carlo updates, is the same that underlies out-of-equilibrium simulations based on Jarzynski's equality, which have been recently deployed to compute free-energy differences in lattice gauge theories. We lay out a strategy to optimize the efficiency of this extended class of generative models and present examples of applications.
\end{abstract}

\end{titlepage}

\section{Introduction}
\label{sec:introduction}

The free energy $F$ is a quantity of central relevance in the description of physical, chemical, or biological systems with a very large (possibly infinite) number of degrees of freedom. For a system in thermal equilibrium at a temperature $T$, its definition\footnote{Throughout this article we work in natural units: $\hbar=c=\kB=1$.} through the equality $F= -T \ln Z$ provides a direct connection between the microscopic physics encoded in the sum of states in the partition function $Z$ and the equation of state describing the macroscopic properties of the system, since in the thermodynamic limit the pressure $p$ equals minus the free-energy density per unit volume. In addition to its importance for low-energy physics systems that traditionally fall into the domain of thermodynamics, the free energy also has applications in high-energy physics, including in elementary particle physics and cosmology. For example, the existence and the nature of phase transitions at very high energies may have an impact on various aspects relevant for the evolution of the early Universe~\cite{Schwarz:2003du, Caprini:2019egz}. While the Standard Model predicts smooth cross-overs both in the electro-weak- and in the strong-interaction sector~\cite{Kajantie:1996mn, Kajantie:1995kf, Aoki:2006we, Bazavov:2011nk}, the existence of first-order phase transitions in various types of extensions of the Standard Model may leave an imprint in gravitational waves~\cite{Witten:1984rs, Hogan:1986qda, Kosowsky:1992rz}.

In quantum chromodynamics (QCD) the theoretical determination of the free energy at temperatures of the order of the hadronic scale necessarily relies on numerical methods, which are based on Wilson's lattice regularization~\cite{Wilson:1974sk} and amount to estimating \emph{ratios} of Feynman path integrals by Markov-chain Monte~Carlo calculations---that is, \emph{differences} in free energies. It is well known, however, that typically this involves significant computational costs: commonly used methods to perform such calculations are based on the numerical integration of a derivative with respect to some parameter~\cite{Engels:1990vr} or on reweighting the field configurations of a simulated ensemble to a target ensemble, specified by different parameter values~\cite{Ferrenberg:1988yz}. The former of these methods, however, introduces a systematic uncertainty due to the discretization of the integration interval; the latter, on the other hand, is often hampered by the fact that the overlap between the most typical configurations in the simulated and in the target ensemble becomes exceedingly small in the thermodynamic limit. The challenging nature of this problem continues to motivate a search for alternative techniques to estimate the free energies in Monte~Carlo lattice QCD~\cite{Langfeld:2012ah, Asakawa:2013laa, Langfeld:2014nta, Kitazawa:2016dsl, Giusti:2016iqr, Kanaya:2016rkt, DallaBrida:2017sxr, Hirakida:2018uoy, Iritani:2018idk}, or, more generally, Feynman path integrals in high-dimensional spaces~\cite{Che:2022gzu}. 

A different computational strategy to evaluate free-energy or effective-action differences has been recently pursued in a series of works~\cite{Chatelain:2006pdo, Chatelain:2007ts, Hijar:2007new, Caselle:2016wsw, Caselle:2018kap, Francesconi:2020fgi} that are based on an exact equality in non-equilibrium statistical mechanics discovered by C.~Jarzynski more than 25 years ago~\cite{Jarzynski:1996oqb, Jarzynski:1997ef}. It expresses the free-energy difference between two equilibrium states of a statistical system in terms of the exponential average of the work done on the system to push it out of equilibrium. Jarzynski's equality is part of a series of works that, during the last decade of the past century, studied in depth the connection between deviations from thermodynamic equilibrium and entropy production~\cite{Evans:1993po, Evans:1993em, Gallavotti:1994de, Gallavotti:1995de, Crooks:1997ne, Crooks:1999ep} (and which are reviewed in refs.~\cite{Ritort:2004wf, MariniBettoloMarconi:2008fd}): it extends and generalizes earlier results~\cite{Bochkov:1977gt, Bochkov:1979fd, Bochkov:1981nf} (for a discussion of the connection between those previous works and Jarzynski's equality, see refs.~\cite{Jarzynski:2006cof, Kuzovlev:2011sr}), entailing a number of implications, in particular, for the scrambling of quantum information and quantum chaos~\cite{Esposito:2009zz, DAlessio:2016rwt, Campisi:2016qlj, Halpern:2016zcm, Halpern:2017abm, Mori:2018qjb, Chenu:2018spm}, while it reduces to known identities in particular limits. Among the implications of Jarzynski's equality we mention the fact that, simply using the mathematical properties of convex functions~\cite{Jensen:1906sl}, it is possible to derive from it the \emph{inequality} that expresses the second law of thermodynamics. In the context of Markov Chain Monte Carlo simulations, a closely-related implementation of the same idea is Annealed Importance Sampling (AIS)~\cite{Neal1998, Neal2001}, which has seen widespread use in several research fields; we observe that the connection with Jarzynski's equality was already made explicit in the original AIS paper.

Recently, the explosive growth of machine-learning applications in virtually all fields of human activity has triggered an avalanche of novel implementations of these techniques also in the physical sciences~\cite{Carleo:2019ptp} and in elementary particle physics~\cite{Guest:2018yhq, Radovic:2018dip, Schwartz:2021ftp}, broadly extending their previous domains of usage~\cite{Denby:1987rk, Lonnblad:1990qp, Denby:1992jd, Mankel:2004yv, Aad:2014yva, Ball:2014uwa, Barnard:2016qma, Louppe:2017ipp}. This also includes applications in lattice field theory: as examples of recent works in this area of research, we mention refs.~\cite{Wetzel:2017ooo, Zhou:2018ill, Shanahan:2018vcv, Urban:2018tqv, Yoon:2018krb, Giannetti:2018vif, Li2018, Matsumoto:2019jia, Chernodub:2020nip, Bluecher:2020kxq, Favoni:2020reg, Boyda:2020nfh, Medvidovic:2020vum, Bachtis:2021xoh, Gabrie:2021tlu, Singha:2021nht}, but this list is likely to grow much longer in the next few years, as the lattice community is developing approaches that are expected to make machine-learning techniques part of the standard lattice-QCD toolbox~\cite{Joo:2019byq}. A class of deep generative models called \emph{normalizing flows}~\cite{Tabak2010:deb, Kobyzev:2019nfa, Papamakarios:2021nff} represents one of the most active and interesting developments in this area of research~\cite{Albergo:2019eim, Kanwar:2020xzo, Boyda:2020hsi, Nicoli:2020njz, Albergo:2021vyo, DelDebbio:2021qwf, Albergo:2021bna, Hackett:2021idh, deHaan:2021erb, Finkenrath:2022ogg}. 
Normalizing flows can be thought of as an invertible map between a latent (easy) distribution and the target probability distribution, whose complexity is encoded in the trainable parameters of the neural networks that compose the flow itself. The fact that the configurations sampled using this kind of generative models are statistically independent is a promising feature, as it represents a completely new way to approach a typical problem that plagues Monte~Carlo simulations close to the continuum limit, namely the so-called critical slowing down. On the other hand, training times seem to grow very quickly when approaching the continuum limit of the theory of interest~\cite{DelDebbio:2021qwf} so more work is needed to improve the scalability of training procedures. Interestingly, it is crucial to note that normalizing flows do not simply provide a new sampling technique for lattice field theories, but represent also a natural tool for the determination of the partition function $Z$. Namely, the importance sampling technique introduced in ref.~\cite{Nicoli:2020njz} allows for the direct estimation of the free energy at certain values of the theory parameters, and not just of its difference with respect to another point in parameter space, thus representing a conceptual evolution with respect to the aforementioned Monte~Carlo free-energy estimation techniques.

Normalizing flows and non-equilibrium Monte~Carlo calculations based on Jarzynski's equality are the two main topics that we study in the present work: despite the obvious differences between these two computational approaches, in the following we point out the existence of a direct connection between them, and show how this relation can be fully exploited using a more general class of generative models called \emph{Stochastic Normalizing Flows}~\cite{Wu:2020snf}. Interestingly, the idea of combining deterministic mappings and out-of-equilibrium transformations used in Jarzynski's equality goes back to ref.~\cite{Vaikuntanathan2011}, albeit for fixed, non-trainable mappings.

The structure of our manuscript is the following: after reviewing the basic aspects of Jarzynski's equality, and discussing its practical use for Monte~Carlo integration in section~\ref{sec:Jarzynski_equality}, in section~\ref{sec:connection_with_normalizing_flows} we reformulate the equality in a framework suitable both for normalizing flows and stochastic processes, starting from the formulation of ref.~\cite{Nicoli:2020njz} and reprising previous work on ref.~\cite{Wu:2020snf}. Next, in section~\ref{sec:phi4_results} we present an example of physical application of these flows in the evaluation of the free energy of the $\phi^4$ two-dimensional lattice field theory, comparing the effectiveness of different types of flows. Finally, in section~\ref{sec:conclusions_and_future_prospects} we recapitulate our findings and discuss possible future extensions of this work.

\section{Jarzynski's equality}
\label{sec:Jarzynski_equality}

Consider a statistical-mechanics system, with degrees of freedom $\phi$, whose dynamics depends on a set of parameters collectively denoted as $\eta$ (which can be the couplings appearing in the Hamiltonian $H$, etc.). Jarzynski's equality~\cite{Jarzynski:1996oqb, Jarzynski:1997ef} states that the ratio of the partition functions corresponding to equilibrium states of the system corresponding to two different values of $\eta$, to be denoted as $\etafin$ and $\etain$, is equal to the \emph{exponential average} of the work $W$, in units of the temperature $T$, that is done on the system, when, starting from thermodynamic equilibrium with parameters $\etain$, it is driven out of equilibrium by a change of its coupling from $\etain$ to $\etafin$ during a time interval $[\tin, \tfin]$, according to a protocol $\eta(t)$:
\begin{equation}
\label{Jarzynski_equality}
\frac{Z_{\etafin}}{Z_{\etain}} = \overline{\exp \left( -W/T \right)}.
\end{equation}
The average (denoted by the bar) appearing on the right-hand side of eq.~(\ref{Jarzynski_equality}) is taken over all possible trajectories in the space of configurations that the system can follow, when its parameters are modified according to $\eta(t)$. While the starting configurations of these trajectories are equilibrium ones, this is no longer the case for all $t>\tin$, as during each trajectory the system is driven out of equilibrium and never allowed to relax to equilibrium anymore. As a consequence, eq.~(\ref{Jarzynski_equality}) describes a non-trivial relation between equilibrium quantities (on the left-hand side) and non-equilibrium ones (on the right-hand side). The time $t$ on which the parameters $\eta$ depend can be either real time or Monte~Carlo time. Eq.~(\ref{Jarzynski_equality}) can be proven in different ways. In appendix~\ref{sec:jarzynski_derivation}, we review a ``constructive'' proof, that is closest to the implementation of Monte~Carlo algorithms to compute free-energy differences by evaluating the right-hand side of the equation above. In this case, the time $t$ is discretized and identified with the Monte~Carlo time, while the work $W$ can be written as
\begin{equation}
 W = \sum_{n=0}^{N-1} \left\{ H_{\eta_{n+1}}\left[\phi_n\right] - H_{\eta_n}\left[\phi_n\right] \right\}
\end{equation}
where $\eta_i = \eta(t_i)$.

It may be surprising that the $\eta(t)$ protocol, which describes ``how'' the parameters of the system are let evolve in time, is fixed and arbitrary, and is not averaged over. As the proof in appendix~\ref{sec:jarzynski_derivation} shows, the result of the $\overline{\exp \left( -W/T \right)}$ average appearing on the right-hand side of eq.~(\ref{Jarzynski_equality}) is independent from $\eta$. From the point of view of a Monte~Carlo implementation, however, the choice of $\eta$ has a strong effect on the efficiency with which the algorithm can produce accurate and precise numerical estimates of the $Z_{\etafin}/Z_{\etain}$ ratio with a finite number of trajectories.

To understand the meaning of eq.~(\ref{Jarzynski_equality}), it is interesting to consider it in two particular limits.

In the limit when the $[\tin, \tfin]$ time interval becomes infinitely long, the $\eta$ parameters evolve infinitely slowly, so that the configurations remain arbitrarily close to thermodynamic equilibrium along each trajectory. Then, the evolution of the system is dissipationless and the work along each trajectory is equal to the free-energy difference between the final and the initial statistical ensembles. In that case, the right-hand side of eq.~(\ref{Jarzynski_equality}) trivially reduces to $\exp(-\Delta F/T)$.

In the opposite limit, when the switching process from $\etain$ to $\etafin$ becomes instantaneous, eq.~(\ref{Jarzynski_equality}) can be written as
\begin{equation}
\label{reweighting}
\overline{\exp \left( -W/T \right)} = \sum_{\phi_0} \sum_{\phi_1} \pi_{\etain}\left[ \phi_0 \right] \exp \left( -\frac{H_{\etafin}\left[\phi_0\right] - H_{\etain}\left[\phi_0\right]}{T}\right) P_{\etafin}\left[ \phi_0 \to \phi_1 \right]
\end{equation}
where $\pi_{\etain}$ stands for the probability distribution of the system with $\eta = \etain$.
The sum over $\phi_1$ is trivial and one is left with:
\begin{equation}
\label{reweighting_bis}
\overline{\exp \left( -W/T \right)} = \sum_{\phi_0} \pi_{\etain}\left[ \phi_0 \right] \exp\left( -\frac{H_{\etafin}\left[\phi_0\right] - H_{\etain}\left[\phi_0\right]}{T}\right).
\end{equation}
The sum on the right-hand side of eq.~(\ref{reweighting_bis}) can be interpreted as an expression for $Z_{\etafin}$ (divided by $Z_{\etain}$) as a weighted sum over the configurations that contribute to $Z_{\etain}$, where the weight of each configuration $\phi_0$ is $\exp\left( -\left\{H_{\etafin}\left[\phi_0\right] - H_{\etain}\left[\phi_0\right]\right\}/T\right)$: this means that in this limit Jarzynski's equality~(\ref{Jarzynski_equality}) simply reduces to the equation describing statistical reweighting~\cite{Ferrenberg:1988yz}.

We also wish to point out that Jarzynski's equality~(\ref{Jarzynski_equality}) is closely related to another important result in non-equilibrium statistical mechanics, Crooks' theorem~\cite{Crooks:1997ne, Crooks:1999ep}. The latter states that the ratio between the probability density $\mathcal{P}_{\mbox{\tiny{f}}}(W)$ that a ``forward'' non-equilibrium transformation cost work $W$, and the probability density $\mathcal{P}_{\mbox{\tiny{r}}}(-W)$ that the opposite (``reverse'') transformation cost work $-W$ is given by
\begin{equation}
\label{Crooks_theorem}
\frac{\mathcal{P}_{\mbox{\tiny{f}}}(W)}{\mathcal{P}_{\mbox{\tiny{r}}}(-W)} = \exp \left( - \frac{\Delta F - W}{T} \right) .
\end{equation}
The connection between eq.~(\ref{Crooks_theorem}) and Jarzynski's equality is obvious, as eq.~(\ref{Jarzynski_equality}) can be obtained by multiplying eq.~(\ref{Crooks_theorem}) by $\mathcal{P}_{\mbox{\tiny{r}}}(-W)$ and integrating over $W$. Note that an interesting implication of Crooks' theorem is that the free-energy difference $\Delta F$ is the value of $W$ for which $\mathcal{P}_{\mbox{\tiny{f}}}(W)$ and $\mathcal{P}_{\mbox{\tiny{r}}}(-W)$ are equal.

Finally, we note that the theoretical results presented in this section for a statistical mechanics system are immediately translatable in the language of quantum field theory by substituting $H[\phi]/T$ with the Euclidean action $S[\phi]$ and $W/T$ with the generalized work
\begin{align}
\label{eq:w_stochastic}
 w(\phi_0,\phi_1,\dots , \phi_N)
  &= \sum_{n=0}^{N-1} \left\{ S_{\eta_{n+1}} \left[\phi_n\right] - S_{\eta_n}\left[\phi_n\right] \right\} \\
  &= S_{\eta_{N}} \left[\phi_N\right] - S_{\eta_0} \left[\phi_0\right] - Q(\phi_0,\phi_1,\dots , \phi_N)
\end{align}
where in the second line we introduced the quantity
\begin{equation}
\label{eq:Q_stochastic}
 Q(\phi_0,\phi_1,\dots , \phi_N) = \sum_{n=0}^{N-1} \left\{ S_{\eta_{n+1}} \left[\phi_{n+1}\right] - S_{\eta_{n+1}}\left[\phi_n\right] \right\}
\end{equation}
that is the equivalent of the heat exchanged with the environment during the transformation defined by the protocol $\eta(t)$. At the end of the next section we will see how the definition of eq.~(\ref{eq:w_stochastic}) can be generalized in a framework that includes also normalizing flows.
Now eq.~(\ref{Jarzynski_equality}) can be written as
\begin{align}
\label{eq:Jarzynski2}
\frac{Z_{\etafin}}{Z_{\etain}} 
 &= \langle \exp \left( -w(\phi_0,\phi_1,\dots , \phi_N) \right)  \rangle_{\mbox{\tiny{f}}} \nonumber \\
 &= \int \dd \phi_0  \, \dd \phi_1 \dots \dd \phi_N \, \pi_{\etain}(\phi_0) \, P_{\mbox{\tiny{f}}}[\phi_0,\phi_1,\dots, \phi_N] \, \exp \left( -w(\phi_0,\phi_1,\dots , \phi_N) \right),
\end{align}
where the average over all possible paths $\etain \to \etafin$ has been expressed through the probability $P_{\mbox{\tiny{f}}}$ of going through a given set of configurations $\phi_0 \to \phi_1 \to \dots \to \phi_N$, having used the distribution $\pi_{\etain}(\phi_0) = e^{-S_{\etain}(\phi_0)}/Z_{\etain}$ to sample $\phi_0$. Let us add that it is inside $P_{\mbox{\tiny{f}}}$ where it lies the dependence of the calculation on crucial details of the transformation, such as the protocol $\eta(t)$ or the Monte~Carlo algorithm chosen to update the system in the intermediate steps. In the following, we will refer to a transformation defined by a given protocol $\eta(t)$ as a ``stochastic evolution''.

We end this section by pointing out that the same equality can be used to compute the expectation value of a generic observable $\mathcal{O}$ at $\eta = \etafin$:
\begin{equation}
\label{eq:expectation_value_0}
\langle \mathcal{O} \rangle_{\eta=\etafin} = \frac{\langle \mathcal{O}(\phi_N) \exp(-w(\phi_0,\phi_1,\dots , \phi_N)) \rangle_{\mbox{\tiny{f}}}}{\langle \exp(-w(\phi_0,\phi_1,\dots , \phi_N)) \rangle_{\mbox{\tiny{f}}}},
\end{equation}
whose derivation follows closely the one of eq.~(\ref{eq:Jarzynski2}) in the appendix~\ref{sec:jarzynski_derivation}. 

\section{Connection with normalizing flows}
\label{sec:connection_with_normalizing_flows}

In this section, we first review the basics about normalizing flows (mostly following the presentation in ref.~\cite{Wu:2020snf}), before exposing the relation between normalizing flows and non-equilibrium Monte~Carlo simulations based on Jarzynski's equality.

Normalizing flows~\cite{Tabak2010:deb, JimenezRezende:2015viw, Kobyzev:2019nfa, Papamakarios:2021nff} can be interpreted as (a discrete collection of) bijective and differentiable functions interpolating between two different statistical distributions,\footnote{More precisely, the differentiability must hold at least almost everywhere in the measurable spaces on which the distributions are defined.} and provide a natural tool to construct scalable, arbitrarily complex approximations of unknown posterior distributions in variational-inference problems. Starting from functions that map a base (or ``prior'') distribution, which is sufficiently simple to be mathematically tractable, to a target distribution, the density of a statistical sample from the latter can be obtained by constructing its counter-image, and multiplying its density by the product of the Jacobians encoding the volume change along the transformation, i.e. the target distribution is the push-forward of the base distribution. The function from the base distribution to the target distribution can be described as a ``generative'' map, as it transforms ``noise'' into the feature-rich, physical target distribution. Conversely, the inverse function is a ``normalizing'' one, mapping the target distribution into the simpler base distribution.

Normalizing flows can be implemented as neural networks by ``discretizing'' the functions that interpolate between the base distribution $q_0$ and the target distribution $p$ through a composition of invertible layers, labeled by a natural number $0 \le n \le N$. If $z$ denotes a variable from the base distribution and $g_\theta$ is the generative map, one can write:
\begin{equation}
\label{eq:norm_flow}
g_\theta (z) = (g_N \circ \cdots \circ g_1 \circ g_0 )(z).
\end{equation}
We denote the distributions of the intermediate variables $y_{n+1}=g_n(y_n)$ as
\begin{equation}
\label{eq:intermediate_dist}
q_{n+1}(y_{n+1})=q_n \left( g_n (y_n) \right) = q_n (y_n )\left| \det J_n (y_n )\right|^{-1},
\end{equation}
where $J_n$ denotes the Jacobian matrix associated with the change of variables between the layers with labels $n$ and $n+1$. The training of the network can be done by minimizing the Kullback--Leibler (KL) divergence between the generated distribution and the target distribution~\cite{Kullback:1951zyt}, which is a measure of the similarity between the two probability distributions and can be written as
\begin{equation}
\label{deterministic_Kullback-Leibler_divergence}
\DKL(q_N\|p) = \int \dd \phi \, q_N(\phi) \left[ \ln q_N(\phi) - \ln p(\phi)\right].
\end{equation}
In a similar fashion with respect to ref.~\cite{Nicoli:2020njz}, we introduce a weight function
\begin{equation}
\label{eq:tilde_w}
\tilde{w}(\phi)=\frac{\exp(-S[\phi])}{Z_0 q_N(\phi)},
\end{equation}
where we included the normalization constant of the $q_0$ distribution in the denominator for reasons that will be clear in the following. The partition function associated with the target probability distribution $p(\phi)$ can be simply expressed as
\begin{equation}
\label{eq:Z_neural}
Z = Z_0 \int \dd \phi \, q_N(\phi) \tilde{w}(\phi) = Z_0 \langle \tilde{w}(\phi) \rangle_{\phi \sim q_N},
\end{equation}
where $\langle \dots \rangle_{\phi \sim q_N}$ denotes the average over the ensemble described by the probability density distribution $q_N$. Then, we are able to write the expectation value of a generic observable $\mathcal{O}$ as
\begin{equation}
\label{expectation_value}
\langle \mathcal{O} \rangle = \frac{1}{Z} \int \dd \phi \, q_N(\phi) \mathcal{O}(\phi) \tilde{w}(\phi) = \frac{\langle \mathcal{O}(\phi) \tilde{w}(\phi) \rangle_{\phi \sim q_N}}{\langle \tilde{w}(\phi) \rangle_{\phi \sim q_N}}.
\end{equation}
Note that the right-hand side of eq.~(\ref{expectation_value}) expresses the expectation value of $\mathcal{O}$ in the target ensemble through a reweighting from the $q_N$ ensemble, which, in turn, is obtained combining the sampling from the base distribution with a deterministic flow $g_\theta$. In particular, the weight function can be rewritten (as a function of $y_0$) in the form 
\begin{align}
\tilde{w}(y_0) 
&= \exp\left( - \left\{ S[g_\theta(y_0)] + \ln Z_0 + \ln q_0[y_0] - Q\right\} \right) \nonumber \\
&= \exp\left( - \left\{ S[g_\theta(y_0)] - S_0[y_0] - Q\right\} \right),
\label{eq:tilde_w_2}
\end{align}
where in the second equation we inserted $q_0[y_0] = \exp\left(-S_0[y_0]\right)/Z_0$. We also defined the quantity $Q$, which encodes the variation in phase-space volume accumulated along the flow:
\begin{equation}
\label{eq:Q_deterministic}
Q=\sum_{n=0}^{N-1} \ln \left| \det J_n (y_n) \right| .
\end{equation}
In practical implementations, $Q$ depends on the network architecture: for example, it is identically zero in frameworks like NICE~\cite{Dinh:2014nni}, while it is generally non-zero for networks based on Real~NVP~\cite{Dinh:2016deu}. Let us note also that eq.~(\ref{expectation_value}) is not the only way to compute expectation values with normalizing flows: a popular alternative consists in generating the configurations using $q_N(\phi)$ and applying an independent Metropolis--Hastings algorithm to correct for the difference between $q_N(\phi)$ and $p(\phi)$~\cite{Tierney1994, Albergo:2019eim}. The acceptance rate of the Metropolis step provides a measure of the quality of the flow.\footnote{In this work we do not pursue this method, but general considerations on the effectiveness of stochastic normalizing flows presented in the following are valid independently of the formula used to compute expectation values.}

Finally, we obtain for eq.~(\ref{eq:Z_neural}) the following form
\begin{equation}
\label{eq:Z_neural_2}
 \frac{Z}{Z_0} = \langle \exp \left( - \left\{ S[g_\theta(y_0)] - S_0[y_0] - Q \right\} \right) \rangle_{y_0 \sim q_0}.
\end{equation}

The reader may have noticed the strong similarities between eq.~(\ref{eq:Z_neural_2}) and eq.~(\ref{eq:Jarzynski2}), in particular when $S_{\etain} $ is identified with $S_0$: this symmetry is not obvious, as the flows used in these computations are purely deterministic in the one case and stochastic in the other. The aim of this section is to generalize both concepts under a common framework, that can describe both deterministic and stochastic transformations and eventually use them together in a single flow.

Following the work of ref.~\cite{Wu:2020snf}, we start by considering a configuration $y_0$ sampled from the base distribution and we define a forward path as a sequence of configurations $(y_0,y_1,\dots,y_t)$, with $t\le N$. The probability of going through this path can be expressed as the product of the transition probabilities at all intermediate steps:
\begin{equation}
P_{\mbox{\tiny{f}}}[y_0,y_1,\dots,y_t]=\prod_{n=0}^{t-1} P[y_n \to y_{n+1}],
\end{equation}
so that the probability of reaching a given configuration $y_t$ at a generic step $t$ can be expressed by integrating over the initial configuration $y_0$ (sampled from the base distribution $q_0$) and over all intermediate configurations:
\begin{equation}
\label{reach_yn_in_stochastic_normalizing_flow}
q_t(y_t)=\int \dd y_0  \, \dd y_1 \dots \dd y_{t-1} \, q_0(y_0) P_{\mbox{\tiny{f}}}[y_0,y_1,\dots,y_t].
\end{equation}
Another useful quantity is the probability of going through the reverse path $(y_t,y_{t-1},\dots,y_0)$
\begin{equation}
P_{\mbox{\tiny{r}}}[y_t,y_{t-1},\dots,y_0]=\prod_{n=0}^{t-1} P[y_{t-n} \to y_{t-n-1}],
\end{equation}
that allows for the definition of the weight function
\begin{align}
\label{eq:tilde_w_general}
\tilde{w}(y_0,y_1,\dots, y_N)
 &=\frac{Z}{Z_0} \frac{p(y_N) P_{\mbox{\tiny{r}}}[y_N,y_{N-1},\dots, y_0]}{q_0(y_0) P_{\mbox{\tiny{f}}}[y_0,y_1,\dots, y_N]} \\
 &= \exp\left( - \left\{ S[y_N]- S_0[y_0] - Q\right\} \right)
\end{align}
with 
\begin{equation}
\label{eq:Q_general_definition}
Q(y_0, y_1, \dots y_N) = \ln \frac{P_{\mbox{\tiny{r}}}[y_N,y_{N-1},\dots,y_0]}{P_{\mbox{\tiny{f}}}[y_0,y_1,\dots,y_N]} =\sum_{n=0}^{N-1} \left( \ln P[y_{n+1} \to y_n] -\ln P[y_n \to y_{n+1}] \right).
\end{equation}
Note that, in our discussion above, the quantity $Q$ defined in eq.~(\ref{eq:Q_deterministic}) was a function of $y_0$ only---a consequence of the deterministic nature of the flow. For flows containing stochastic steps, $Q$ depends on all $y_n$, for $0 \le n \le N$, i.e. is a function of a particular trajectory, not only of its starting (or of its final) point. The KL~divergence can then be interpreted as a ``distance'' between the forward and reverse paths that go through the same configurations\footnote{Strictly speaking, eq.~(\ref{eq:general_Kullback-Leibler_divergence}) does not define an actual metric; in particular, it does not necessarily satisfy the triangle inequality.}, i.e.
\begin{align}
\label{eq:general_Kullback-Leibler_divergence}
\DKL(q_0 P_{\mbox{\tiny{f}}} \| p P_{\mbox{\tiny{r}}}) 
&= \int \dd y_0  \, \dd y_1 \dots \dd y_N \, q_0 (y_0) P_{\mbox{\tiny{f}}}[y_0,y_1,\dots, y_N] \ln \frac{q_0(y_0) P_{\mbox{\tiny{f}}}[y_0,y_1,\dots, y_N]}{p(y_N) P_{\mbox{\tiny{r}}}[y_N,y_{N-1},\dots, y_0]} \nonumber \\
&= - \langle \ln \tilde{w}(y_0,y_1,\dots, y_N) \rangle_{\mbox{\tiny{f}}} + \ln \frac{Z}{Z_0}
\end{align}
while the ratio of partition functions simply becomes
\begin{equation}
\label{eq:Z_final}
\frac{Z}{Z_0} = \langle \tilde{w}(y_0,y_1,\dots, y_N) \rangle_{\mbox{\tiny{f}}}.
\end{equation}

Both normalizing flows and the stochastic procedure described in section~\ref{sec:Jarzynski_equality} emerge naturally within this framework. Normalizing flows are easily recovered by setting 
\begin{equation}
P[y_n \to y_{n+1}]=\delta\left(y_{n+1}-g_n(y_n)\right)
\end{equation}
for every $n$, with $g_n(y_n)$ being the transformation of layer $n$. In that case, using the fact that
\begin{equation}
\label{Bayes_theorem}
q_n(y_n)P[y_n \to y_{n+1}] = q_{n+1}(y_{n+1})P[y_{n+1} \to y_n],
\end{equation}
the right-hand side of eq.~(\ref{eq:Q_general_definition}) reduces to 
\begin{equation}
\sum_{n=0}^{N-1} \ln \left[ q_n(y_n)/q_{n+1}(y_{n+1})\right] =\sum_{n=0}^{N-1} \ln \left| \det J_n (y_n) \right|,
\end{equation}
i.e. to eq.~(\ref{eq:Q_deterministic}). Similarly, the quantity $P_{\mbox{\tiny{f}}}[y_0,y_1,\dots, y_N]$ appearing in eq.~(\ref{eq:general_Kullback-Leibler_divergence}) reduces to a product of Dirac distributions, and after integration this definition of $\DKL(q_0 P_{\mbox{\tiny{f}}} \| p P_{\mbox{\tiny{r}}})$ leads to eq.~(\ref{deterministic_Kullback-Leibler_divergence}).

For the stochastic procedure used in Jarzynski's equality, we first introduce the protocol $\eta(t)$ to interpolate between the base distribution and the target distribution. As we remarked above, the $\eta(t)$ function (or its discretization on the layers specified by an integer-valued label) is largely arbitrary, provided it satisfies the requirements of yielding the parameters of the base and target distributions at the initial and final times, respectively. Given a (discretized) protocol $\eta$, one can construct the sequence of Boltzmann distributions at the $n$-th step: 
\begin{equation}
\label{Boltzmann_distribution_for_lambda_n}
\pi_n[\phi] = \pi_{\eta_n}[\phi] = \frac{1}{Z_{\eta_n}} \exp \left( - S_{\eta_n}[\phi]\right)
\end{equation}
and construct transition probabilities $P(\phi \to \phi^\prime)$ satisfying detailed balance (a condition analogous to eq.~(\ref{Bayes_theorem})):
\begin{equation}
\label{detailed_balance_for_lambda_n}
\frac{P[\phi\to\phi^\prime]}{P[\phi^\prime \to \phi]} = \frac{\pi_{n+1}[\phi^\prime]}{\pi_{n+1}[\phi]}.
\end{equation}
Using eq.~(\ref{detailed_balance_for_lambda_n}), eq.~(\ref{eq:Q_general_definition}) can be rewritten in the form
\begin{equation}
Q(y_0, y_1, \dots y_N) = \sum_{n=0}^{N-1} \ln \frac{\pi_{n+1}[y_n]}{\pi_{n+1}[y_{n+1}]},
\end{equation}
which, when combined with eq.~(\ref{Boltzmann_distribution_for_lambda_n}) brings us back to eq.~(\ref{eq:Q_stochastic}). Similarly, the weight $\tilde{w}$ defined in eq.~(\ref{eq:tilde_w_general}) can be rewritten as
\begin{equation}
\label{eq:tilde_w_stochastic}
\tilde{w}(y_0,y_1,\dots , y_N) = \exp \left( - \sum_{n=0}^{N-1} S_{\eta_{n+1}}[y_n]-S_{\eta_{n}}[y_n] \right) = \exp \left( -w(y_0,y_1,\dots , y_N) \right),
\end{equation}
where in the argument of the exponential we have recognized the work $w$ done on the system during the $(y_0,y_1,\dots , y_N)$ trajectory, as defined in eq.~(\ref{eq:w_stochastic}). Thus, eq.~(\ref{eq:Z_final}) is just Jarzynski's equality~(\ref{Jarzynski_equality}) in a form that can be easily translated both in the language of (deterministic) normalizing flows and of Markov~Chain Monte~Carlo simulations.

It is clear now that nothing prevents us from creating a ``stochastic'' normalizing flow that contains both deterministic coupling layers and stochastic updates. In the following section we review some possible applications and advantages of such a choice.

\section{Application in lattice $\phi^4$ field theory}
\label{sec:phi4_results}

We have performed a series of tests in the two-dimensional $\phi^4$ interacting field theory defined on a lattice $\Lambda$ of size $L_t \times L_s$, with lattice spacing $a$. We denote with $N_t=L_t/a$ and $N_s=L_s/a$ the number of sites in the temporal and spatial directions respectively and we impose periodic boundary conditions along both of them. The Euclidean action of the theory is defined as
\begin{equation}
\label{eq:phi4_action}
 S(\phi) = \sum_{x \in \Lambda} - 2 \kappa \sum_{\mu=0,1} \phi(x)\phi(x+\hat{\mu}) + (1 - 2\lambda) \phi(x)^2 + \lambda \phi(x)^4
\end{equation}
and the (target) probability distribution is
\begin{equation}
 p(\phi) = \frac{1}{Z} e^{-S(\phi)},
\end{equation}
where $Z$ denotes the partition function:
\begin{equation}
 Z = \int \prod_{x \in \Lambda} \dd \phi(x) e^{-S(\phi)}.
\end{equation}

We use three different approaches to generate asymptotically correct configurations: purely stochastic protocols (as described in section~\ref{sec:Jarzynski_equality}), standard normalizing flows and stochastic normalizing flows (SNF), in which the affine layers that compose a typical normalizing flow are combined with stochastic layers where Monte~Carlo updates are performed.

In each kind of flow, we sample the latent variables $z$ from a normal distribution
\begin{equation}
 q_0(z) =  \left (\frac{1}{\sqrt{2 \pi \sigma^2}} \right)^{\left| \Lambda \right|} e^{-S_0(z)}
\end{equation}
with a Gaussian action
\begin{equation}
\label{eq:gauss_action}
 S_0(z) = \sum_{x \in \Lambda} \frac{z(x)^2}{2 \sigma^2}.
\end{equation}
We set $\sigma=0.5$ so that we exactly recover eq.~(\ref{eq:phi4_action}) with $\kappa=0$ and $\lambda=0$. This simplifies the protocol that is needed for purely stochastic evolutions when interpolating between $q_0(z)$ and $p(\phi)$.

The main observable of interest is the free-energy density of the system $f= F/L_s = F / (aN_s)$. Since
\begin{equation}
F = - T \ln Z = - \frac{1}{N_t a} \ln Z
\end{equation}
we can look at the dimensionless quantity
\begin{equation}
\label{eq:free_energy_density}
a^2 f = - \frac{\ln Z}{N_t N_s}.
\end{equation}

Recalling the definition of a normalizing flow in eq.~(\ref{eq:norm_flow}), we use as building blocks of the flow the coupling layers $g_i$. In order to ensure invertibility and an easy evaluation of the Jacobian, we define the coupling layers by splitting the lattice into two different partitions. A given layer $g_i$ leaves one partition unchanged while acting on the other one. More precisely, we use an even-odd (or ``checkerboard'') partitioning, so that each subsequent configuration $y^{i+1} = g_i(y^{i})$ can be written as
\begin{align}
\label{eq:affine_layer}
g_i :
\begin{cases}
 y^{i+1}_{\mbox{\tiny{A}}} = y^{i}_{\mbox{\tiny{A}}} \\
 y^{i+1}_{\mbox{\tiny{B}}} = e^{-s(y^{i}_{\mbox{\tiny{A}}})} y^{i}_{\mbox{\tiny{B}}} + t(y^{i}_{\mbox{\tiny{A}}})
 \end{cases}
\end{align}
so that even sites are left unchanged when $\mbox{A} = \mbox{even}$ and $\mbox{B} = \mbox{odd}$, and vice versa for odd sites. We observe that $s$ and $t$ are two different neural networks that take as input a configuration and release as output an equally sized configuration. This setup is commonly referred to as an affine layer and is part of the Real~NVP architecture~\cite{Dinh:2016deu}. 

When building a normalizing flow, it is a desirable feature for it to be equivariant under the symmetries of the probability distribution it is going to approximate. In this case, the target probability distribution $p(\phi)$ is invariant under $\phi \to -\phi$ transformations: to enforce this $\mathbb{Z}_2$ symmetry in the flow we require the mapping $g_\theta$ to be an odd function with respect to $z$. We do so by choosing neural networks with a hyperbolic-tangent activation function for both $s$ and $t$ in eq.~(\ref{eq:affine_layer}), and also by taking the absolute value of the output of $s$. The resulting distribution $q_N (\phi)$ is then invariant under $z \to -z$ transformations. 

The networks used in this work are shallow, i.e. with a single hidden layer between input and output. We obtained results with two types of networks: fully connected networks with $N_s \times N_t$ neurons in the hidden layer and convolutional networks with kernel size $3\times 3$ and one feature map.

The stochastic evolutions described in section~\ref{sec:Jarzynski_equality} can be thought of as a composition of subsequent ``layers'' as well, whose structure can be written very similarly to that of affine layers. In this case, the $n$-th layer is defined by the protocol parameters $\eta_n=\eta(t_n)$ that are used to update the system with the action $S_{\eta_n}$. Exploiting the locality of the action~(\ref{eq:phi4_action}), we perform an even-odd partitioning: a stochastic layer acts on an intermediate configuration by updating odd (or even) sites using an algorithm that uses as input only even (or odd) sites, which in turn are kept fixed, in a similar fashion as the affine layer of eq.~(\ref{eq:affine_layer}). While in this work a highly-efficient heatbath algorithm customized for the target distribution of the $\phi^4$ action has been used, we observe that the Metropolis--Hastings algorithm can be used in this approach as well.

Finally, a protocol $\eta(t)$ has to be set in order to interpolate between the initial and the final action, in this case eqs.~(\ref{eq:gauss_action}) and (\ref{eq:phi4_action}) respectively. In practice, one has to gradually change the values of the parameters of the theory to interpolate from the prior distribution ($\kappa=0$ and $\lambda=0$) to the target distribution at the desired values of $\kappa$ and $\lambda$. In this work we always followed a linear protocol in all parameters and for each layer we applied only one heatbath update. We stress, however, that different protocols, such as non-linear ones or with multiple Markov-chain Monte~Carlo updates in the same layer, are possible, as well as the possibility to let the intermediate parameters be tunable whenever the flow undergoes a training procedure. 

Having fixed the details of affine and stochastic parameters, any flow used in this work is simply characterized by the number of stochastic and deterministic layers: using even-odd partitioning in both cases, we have always used ``blocks'' of two subsequent layers where both even and odd sites are updated once. In the following we denote the number of affine blocks as $\nab$ and the number of stochastic blocks as $\nsb$. In the case of stochastic normalizing flows, where both $\nab \neq 0$ and $\nsb \neq 0$, stochastic blocks are always inserted equally distanced between affine blocks so to maximize the number of deterministic layers between them. For example, in the case $\nab = 2 \nsb$, the flow is built alternating two affine blocks and one stochastic block.

In the case of normalizing flows and SNFs, we perform the training procedure needed to tune the parameters of the neural networks contained in the affine layers of eq.~(\ref{eq:affine_layer}) by minimizing the loss function $-\langle \ln \tilde{w} \rangle_{\mbox{\tiny{f}}}$, which equals the KL divergence (\ref{eq:general_Kullback-Leibler_divergence}) minus the ratio $Z/Z_0$. In order to evaluate the convergence of the training (i.e. the ability of the latent distribution $q_N(\phi)$ to describe the target distribution $p(\phi)$), we also monitor the variance of the loss
\begin{equation}
\mbox{Var}_{\mbox{\tiny{f}}} (\mathcal{L}) = \mbox{Var}_{\mbox{\tiny{f}}} (-\ln \tilde{w}), 
\end{equation}
and the effective sample size (ESS)
\begin{equation}
 \label{eq:ESS}
\mbox{ESS} = \frac{\langle \tilde{w}\rangle_{\mbox{\tiny{f}}}^2}{\langle \tilde{w}^2 \rangle_{\mbox{\tiny{f}}}}
\end{equation}
which is always in the range $[0,1]$ and tends to $1$ for a ``perfect'' training. These quantities can be calculated for a stochastic flow as well, using the same definitions, and determine the quality of the protocol chosen for the flow. 

A few comments are in order regarding the training of SNFs and the Monte~Carlo update algorithm of choice. In this work, the gradients of the tunable parameters of the neural networks, which are computed during the training procedure using the backpropagation algorithm, are propagated also through the stochastic blocks: thus, Monte~Carlo updates are performed also during the training. In general, the question whether the graph used for the computation of the gradients is continuous or not through the stochastic blocks can be answered only by looking at the specific update algorithm. The accept-reject step is a non-differentiable function, so the gradients cannot propagate through it: however the graph will depend on the type of proposal used to generate the updated variable. In the case of the heatbath algorithm, the computational graph of the gradients is generated in a non-trivial manner, as every new (accepted) variable is proposed using the values of the variables on the nearest-neighbour sites.

The training procedure for normalizing flows and SNFs was performed by applying $10^4$ steps ($5 \times 10^4$ for fully-connected architectures) of the ADAM algorithm~\cite{Kingma2014} (with batches of 8000 configurations) to update the parameters of the neural networks; the code used for the training is based on the PyTorch library. We used the ReduceLROnPlateau scheduler with an initial learning rate set to 0.0005 and a patience of 500 steps. All numerical results for the free energy density $a^2 f$ defined in eq.~(\ref{eq:free_energy_density}) are obtained always taking $\nmeas=2\times 10^5$ independent measurements for the average of eq.~(\ref{eq:Z_final}), for any of the three methods used in this work (normalizing flows, SNFs and stochastic evolutions); at the same time also the ESS is calculated using eq.~(\ref{eq:ESS}). The errors on these two quantities have been computed using a jackknife procedure. All the numerical experiments discussed in this section (both training procedures and measurements) have been performed on a NVIDIA Volta V100 GPU with 16GB of memory.  

\subsection{Results for stochastic evolutions}

\begin{figure}[!htb]
\begin{center}
\includegraphics*[width=\textwidth]{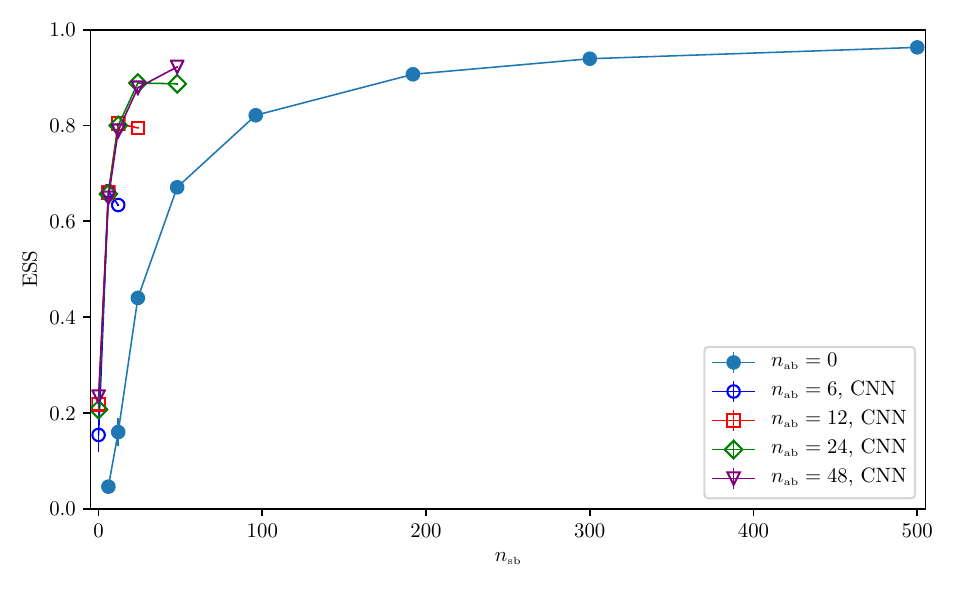}
\caption{Effective sample size for different flows, for varying number of affine blocks $\nab$ and stochastic blocks $\nsb$, for $16\times8$ lattices at $\kappa=0.2$, $\lambda=0.022$. The $\nab=0$ points are purely stochastic evolutions, where no training is required, while the $\nsb=0$ data represent standard normalizing flows. The remaining points in the plot represent SNFs, where stochastic blocks are placed between affine blocks. Error bars are not visible for most of the flows and stochastic evolutions under consideration due to the very small statistical errors.\label{fig:ESS}}
\end{center}
\end{figure}

\begin{figure}[!htb]
\begin{center}
\includegraphics*[width=\textwidth]{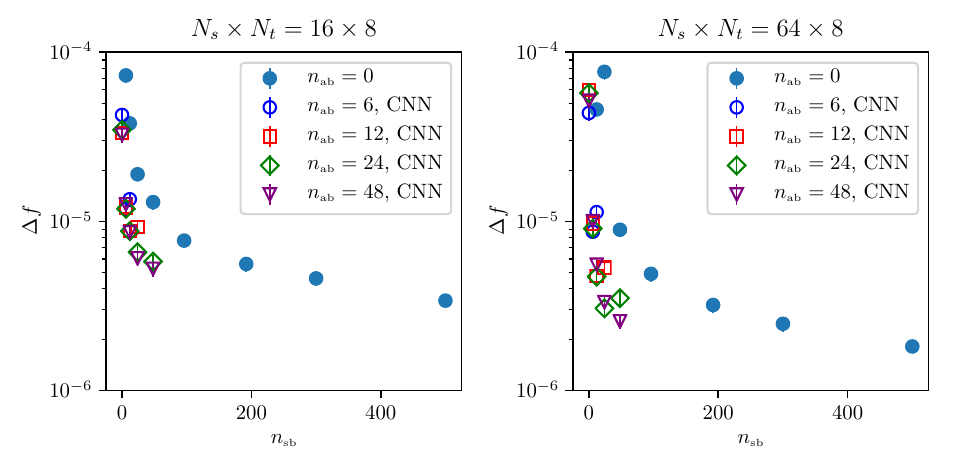}
\caption{\label{fig:f_error} Error on the free energy density $a^2 f$ for various flow architectures from $2 \times 10^5$ independent measurements. Results for $16\times8$ lattices (left-hand-side panel) and $64 \times 8$ lattices (right-hand-side panel) at $\kappa=0.2$, $\lambda=0.022$. The error on the error is calculated with a jackknife procedure.}
\end{center}
\end{figure}

Let us first describe our results for the free-energy density with stochastic evolutions, obtained by computing the average of eq.~(\ref{eq:tilde_w_stochastic}) with $\nmeas$ independent measurements. We stress that in this case no training is strictly needed, as we already fixed all the parameters by choosing a linear protocol.

In figure~\ref{fig:ESS} we report some values obtained for the ESS~(\ref{eq:ESS}) by performing measurements for different flows. The only difference between the various protocols is the number of intermediate steps (or, equivalently, the number of stochastic blocks $\nsb$) between the initial and final points. An effective strategy with this kind of out-of-equilibrium transformations is to increase the number of intermediate Monte~Carlo updates performed during the transformation while keeping the protocol $\eta(t)$ and the number of measurements $\nmeas$ fixed. In this way, each measurement becomes more expensive from a computational point of view, but the distribution $q_N(\phi)$ is also more effective at describing the target distribution $p(\phi)$, as shown by the very high values of ESS obtained for the largest values of $\nsb$ and by the steadily decreasing error in both panels of fig.~\ref{fig:f_error}.

\begin{figure}[!htb]
\begin{center}
\includegraphics*[width=\textwidth]{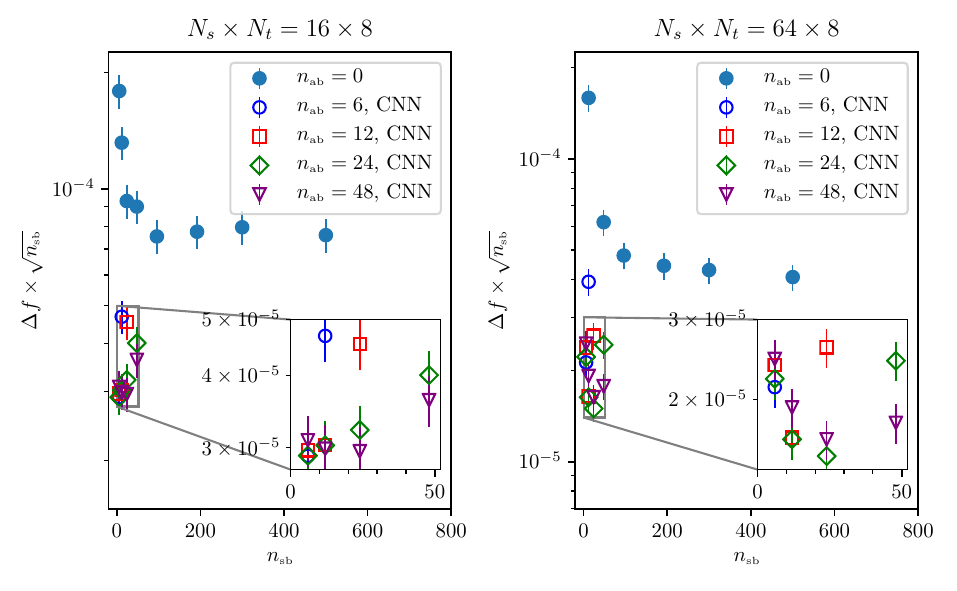}
\caption{\label{fig:f_efficiency} Efficiency of various SNF architectures, determined as the error $\Delta f$ times the square root of the number of stochastic blocks, for $16\times8$ lattices (left-hand-side panel) and $64 \times 8$ lattices (right-hand-side panel). Training time and measurements cost for the deterministic layers in SNFs are not taken into account.}
\end{center}
\end{figure}

To determine the optimal number of stochastic blocks at fixed computational effort, we combined the error on the free-energy density $f$ and an estimate of the computational cost of each measurement. In this case the latter is simply given by $\nsb$, as just a single Monte~Carlo update is performed in each layer.\footnote{Increasing the number of updates in each layer is possible, but we found it not to be helpful in this setup.} The results of this comparison are shown in fig.~\ref{fig:f_efficiency}: one can observe that on the smaller lattice increasing the number of intermediate steps above $100$ does not appear to be particularly cost-effective in sampling the target distribution, while it is still slightly advantageous for the larger volume.

\subsection{Including stochastic layers in normalizing flows}

\begin{figure}[!htb]
\begin{center}
\includegraphics*[width=\textwidth]{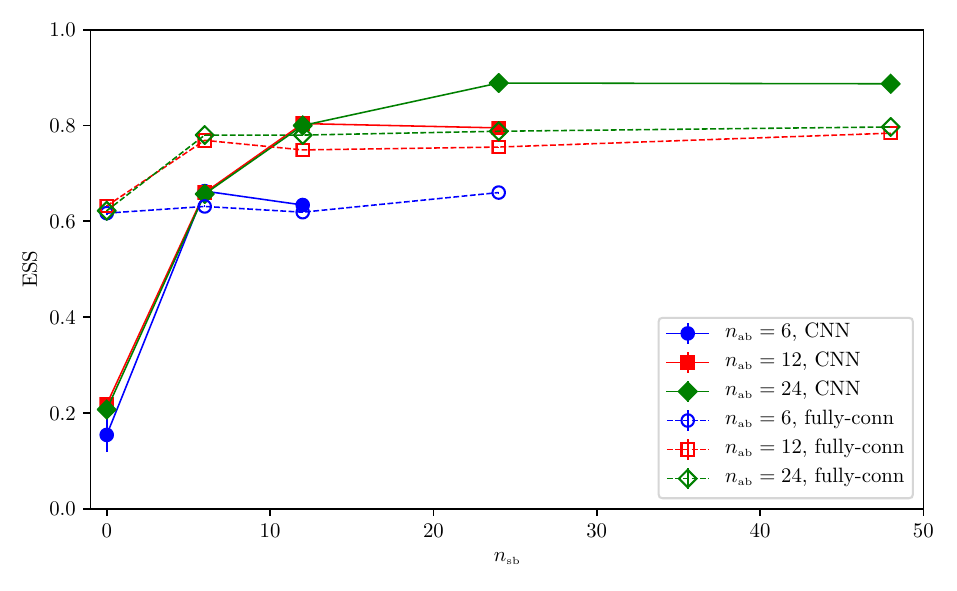}
\caption{Comparison of ESS for stochastic normalizing flows, with fully connected neural networks (empty symbols) and with CNN (full symbols), for varying number of affine blocks $\nab$ and stochastic blocks $\nsb$. Results obtained for $16\times8$ lattices at $\kappa=0.2$, $\lambda=0.022$. Training duration is $10^4$ epochs for SNFs with CNN and $5\times 10^4$ for SNFs with fully connected networks. \label{fig:ESS_cnn_mlp}}
\end{center}
\end{figure}

Let us proceed to the analysis of results for stochastic normalizing flows, where stochastic blocks are inserted between ``deterministic'' affine blocks. Firstly, we observe that for standard normalizing flows, a larger $\nab$ does not necessarily provide a more efficient way of sampling $p(\phi)$, possibly because of a more difficult training. This occurs both for convolutional and fully connected neural networks, as shown in fig.~\ref{fig:ESS_cnn_mlp} for the data at $\nsb=0$: we also add that with the shallow representations used in this setup, the latter perform much better than the former. 

Whenever stochastic layers are inserted between affine layers, the effectiveness of the flows at fixed number of training steps\footnote{We note that increasing $\nsb$ also increases the time required for each training step.} improves in a decisive manner for flows based on convolutional neural networks (CNN), but only slightly for fully connected networks, so that the latter are quickly surpassed for $\nab>6$ by the convolutional architectures. Interestingly, in both cases flows with larger $\nab$ seem to perform better when also $\nsb$ grows, until a plateau is reached. For fully connected networks this plateau is reached very quickly, as the improvement for $\nab>6$ is very small. On the other hand, for CNNs the performance keeps improving even more when $\nab$ increases, especially for larger volumes: it is also interesting to note that flows with $\nab = \nsb$ seem to be more efficient than flows with $\nab \neq \nsb$. Looking at the two insets of fig.~\ref{fig:f_efficiency}, this behaviour is mostly absent for the $16 \times 8$ lattice, but it can be observed for the largest one ($64\times 8$), at least for $0 < \nab < 48$.

\begin{figure}[!htb]
\begin{center}
\includegraphics*[width=\textwidth]{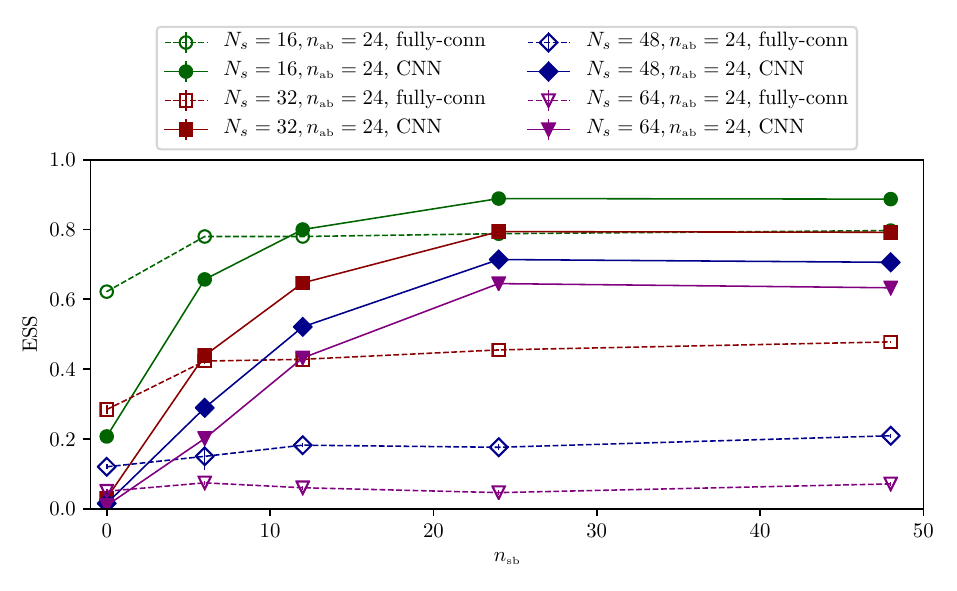}
\caption{Comparison of effective sample size for $N_s \times 8$ lattices between SNFs with fully connected networks (empty points) and CNNs (full points), at $\kappa=0.2$, $\lambda=0.022$. Training duration is the same as in fig.~\ref{fig:ESS_cnn_mlp}. \label{fig:ESS_vol_cnn_mlp}}
\end{center}
\end{figure}

The difference in behavior for the two types of neural networks is even more striking when the spatial size $a N_s$ is increased, as shown in fig.~\ref{fig:ESS_vol_cnn_mlp}. Fully connected networks show little to no improvement when more stochastic blocks are inserted, while CNN-based flows improve with increasing $\nsb$ at a similar rate even for larger volumes.

We stress that the saturation effect that we observe by increasing $\nab$ for SNFs with fully-connected networks in figs.~\ref{fig:ESS_cnn_mlp} and \ref{fig:ESS_vol_cnn_mlp} is not an artifact of training procedures cut too short: we checked that all the combinations of $\nab$ and $\nsb$ under investigation were well in the slowly improving regime and that the ESS had already reached a plateau, both for fully-connected networks and CNNs. As a further check, we performed longer, independent training procedures for the smallest and largest volumes for SNFs with fully-connected networks with $\nab=24$ and $\nsb=6,12,24,48$: for a training twice as long (i.e. after $10^5$ epochs) we obtained values for the ESS around 1\% better than those from shorter trainings.

The neural networks $s$ and $t$ of eq.~(\ref{eq:affine_layer}) used in this work can be considered shallow, since a single hidden layer is present. However, in this setup the overall architectures are ``deep'' in the sense that the number of affine blocks is relatively large ($\nab \geq 6$). What we found through careful experimentation is that flows with large $\nab$ appear to be the easiest to integrate whenever many stochastic layers are inserted. For example, flows with deeper neural networks or less affine blocks were generally less amenable to improvements in training when trained with many stochastic layers.

At this point we would like to use a word of caution concerning the generality of the results obtained in this work. A complete discussion would take into account the dependence on a wide variety of factors: the architecture of affine layers (e.g. the type of neural networks), the characteristics of stochastic layers (e.g. their position with respect to affine layers or the Monte~Carlo algorithm of choice) and the tuning of the hyperparameters. Hardware, too, plays a role: GPUs allow for greater parallelization but, for example, the dependence on the batch size is not obvious when both Markov-chain updates and forward passes on affine coupling layers are performed. For these reasons, a full quantitative comparison between normalizing flows (or SNFs) and purely stochastic evolutions, while extremely interesting for future practical applications, is beyond the scope of this work. Furthermore, including the training time in the overall effort to reach a given error on $f$ is not completely straightforward. Only neural networks strictly need a training procedure to work properly, while for purely stochastic evolutions a reasonably efficient protocol can be set manually. We limit ourselves to observe that in the standard and stochastic normalizing flows used in this work the training time was of the order of hours. This is not negligible at all when compared with the time needed to perform $\nmeas=2\times 10^5$ measurements, which ranges from seconds for flows containing zero or few stochastic layers, to at most a few minutes for flows with $\nsb > 100$.

\subsection{Scaling with the volume}

\begin{figure}[!htb]
\begin{center}
\includegraphics*[width=\textwidth]{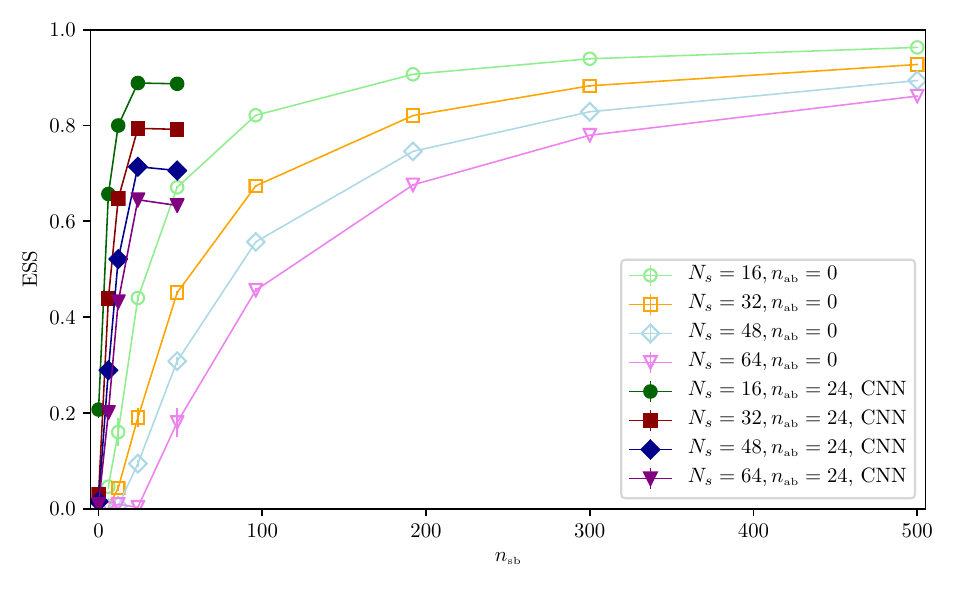}
\caption{Comparison of effective sample size for $N_t = 8$ lattices between purely stochastic evolutions (light colors) and SNFs with CNNs and $\nab=24$ (dark colors) as a function of $\nsb$, at $\kappa=0.2$, $\lambda=0.022$. \label{fig:ESS_vol}}
\end{center}
\end{figure}

\begin{figure}[!htb]
\begin{center}
\includegraphics*[width=\textwidth]{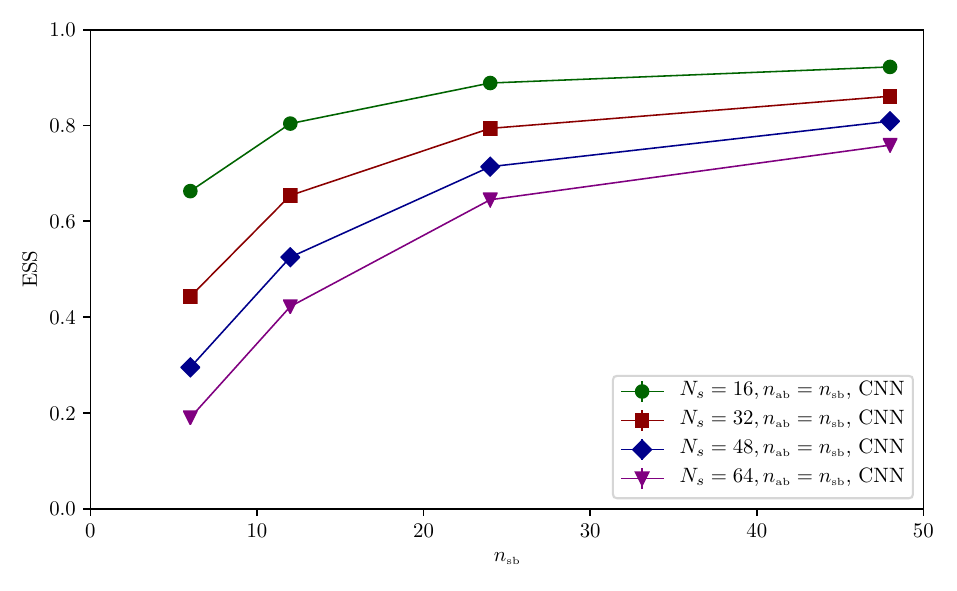}
\caption{ESS comparison for $N_s \times 8$ lattices between CNN-based SNFs with $\nsb=\nab$, at $\kappa=0.2$, $\lambda=0.022$. \label{fig:ESS_vol_2}}
\end{center}
\end{figure}

It is interesting to investigate the effects of an increase in the spatial volume of the lattice: heuristically, the target distribution becomes more sharply peaked and thus a better training and/or a more expressive flow are needed in order to reach the same effectiveness in sampling $p(\phi)$. Before looking at the results, we remind that the number of training steps for SNFs with CNNs is equal to $10^4$ for all the volumes investigated in this work.

For what concerns purely stochastic evolutions, a simple strategy allowing to sample larger volumes as effectively as smaller ones consists in increasing the number of Monte~Carlo updates of the protocol. Intuitively, the corresponding transformation becomes closer and closer to a reversible one, i.e. a transformation in which every intermediate point is (almost) at equilibrium. This can be seen in fig.~\ref{fig:ESS_vol}, where for all volumes the effective sample size grows with $\nsb$.

The question is whether this strategy can be implemented structurally for stochastic normalizing flows as well. As we already pointed out, an improvement in the effectiveness of SNFs can be obtained simply adding deterministic and stochastic blocks in an equal manner, roughly keeping $\nab = \nsb$ and alternating one block of even-odd affine layers with one block of even-odd Monte~Carlo updates. This can be easily seen for all the volumes under study by looking at fig.~\ref{fig:ESS_vol}: for $\nsb = 0$ the ESS is essentially zero except for the smallest volume, but for $\nsb > 0$ it grows quickly for all values of $N_s$ (with $N_t$ left fixed) until reaching the point $\nab = \nsb = 24$; after that, no improvement is observed if $\nsb$ increases. 
In order to further improve a flow then, $\nab$ must be increased as well: when $\nab=\nsb$ the value of the ESS for all volumes steadily increases, as shown in fig.~\ref{fig:ESS_vol_2}.

We would like to note again that these results are obtained when the training of all architectures under consideration was already well inside the slowly-improving regime. Even increasing the training duration by a factor 5 (i.e. to $5\times 10^4$ epochs) the improvement in the quality of the training is absent (for smaller values of $\nab$) or very limited (around 1\% increase in the ESS for $\nab=48$) both for the smallest and largest volumes under study: most importantly, the general behaviour previously observed when increasing $\nsb$ and $\nab$ still holds.

\begin{figure}[!htb]
\begin{center}
\includegraphics*[width=\textwidth]{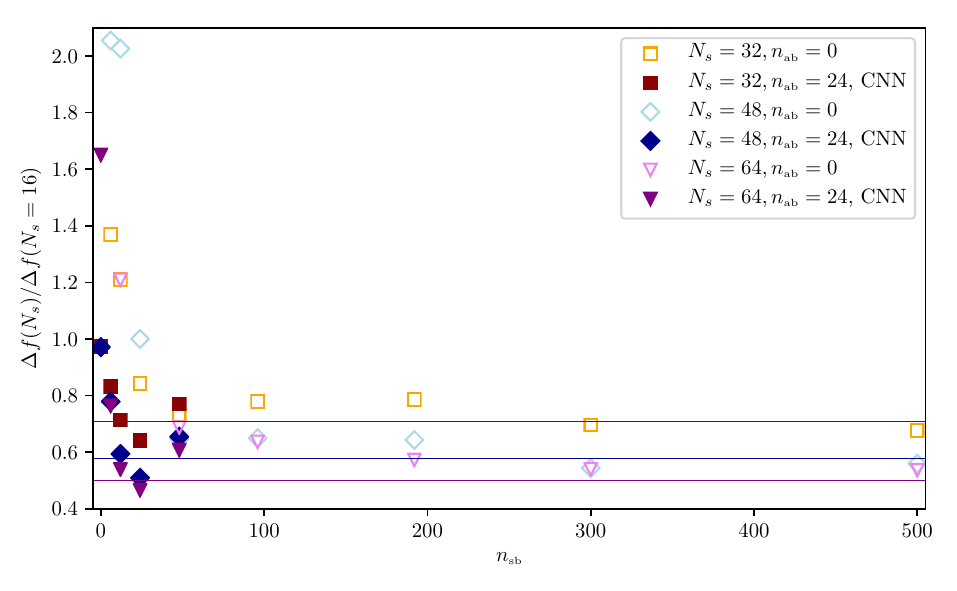}
\caption{Error $\Delta f$ of the free-energy density for $N_s \times 8$ lattices normalized to the error for a $16 \times 8$ lattice, for different flows at $\kappa=0.2$, $\lambda=0.022$ and with $\nmeas = 2 \times 10^5$. Horizontal lines show the error decrease that is expected considering only the volume averaging.\label{fig:error_ratio_vol}}
\end{center}
\end{figure}

Naively, larger lattices allow for smaller errors, as averaging over bigger volumes is akin to having increased statistics. However, as we remarked above, the target distribution $p(\phi)$ is more difficult to sample on larger volumes, so it is interesting to investigate the effort required to obtain the same effectiveness when changing the lattice spatial size. In figure~\ref{fig:error_ratio_vol} we study the error reduction with respect to $N_s=16$, when increasing $\nsb$ either for stochastic evolutions, or for SNFs with $\nab=24$ affine blocks. In the first case, the naive error reduction is reached around $\nsb=300$ for $N_s = 32$ and $48$, while for $N_s=64$ a protocol with $\nsb=500$ might not be sufficient yet. For SNFs the situation is rather different though, as the expected gain is already reached for $\nsb=24$, which, as discussed above, is the most efficient setup for $\nab=24$. This could indicate that for these flows the increase in volume can be compensated more easily than by simply increasing $\nsb$ in purely stochastic evolutions.

\subsection{Exploring the parameter space}

\begin{figure}[!htb]
\begin{center}
\includegraphics*[width=\textwidth]{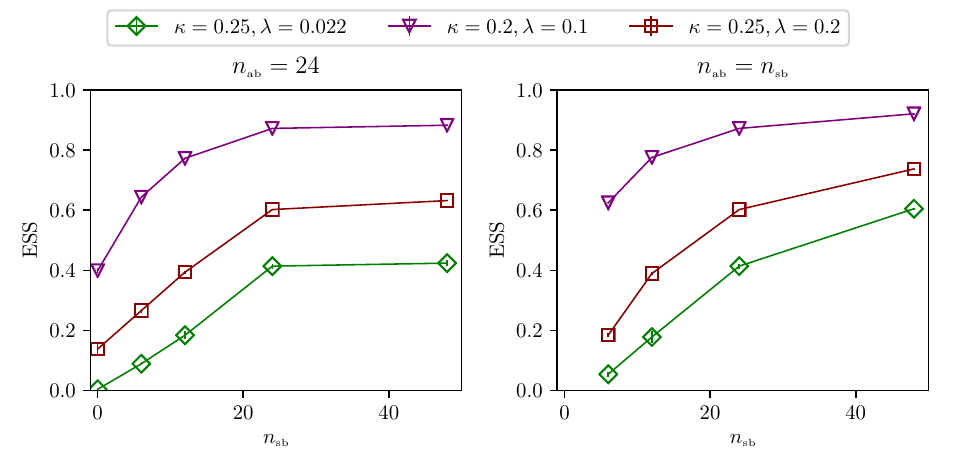}
\caption{ESS comparison for $16 \times 8$ lattices between CNN-based SNFs with $\nsb=24$ (left panel) and $\nsb=\nab$ (right panel), for different values of the target parameters. \label{fig:ESS_otherpars}}
\end{center}
\end{figure}

In order to consolidate the results obtained for $\kappa=0.2$ and $\lambda=0.022$, we explore further the unbroken symmetry phase of the model by changing both target parameters. The trends that we observe in fig.~\ref{fig:ESS_otherpars} are fully compatible with those discussed previously in this section, namely the saturation of the ESS whenever $\nsb > \nab$ and the clear improvement in the quality of the flows if both the number of affine and stochastic blocks are increased while being kept equal.

A word of caution is in order concerning the extrapolation of these results in any point of the parameter space, in particular in the vicinity of the transition to the broken symmetry phase. A different kind of SNF architectures might be needed to reproduce the same trends observed in this section.

\section{Conclusions and future prospects}
\label{sec:conclusions_and_future_prospects}

In this manuscript we have made explicit the connection between normalizing flows and non-equilibrium Monte~Carlo calculations based on Jarzynski's equality (``stochastic evolutions''), which becomes apparent when both are used for the determination of the free energy $F$. In general, stochastic evolutions by themselves represent an efficient method to sample a target distribution and to directly evaluate the partition function $Z$: they provide a novel framework to compute expectation values in lattice field theory from first principles and represent a viable alternative to the traditional Monte~Carlo approach. In Markov-chain Monte~Carlo simulations, measurements are performed on subsequent equilibrium configurations that are part of the same Markov chain, thermalized according to the target coupling(s). The cost of generating a new configuration is as low as the application of a single update on a lattice. However, configurations belonging to the same Markov chain are correlated with each other, reducing the number of effectively independent configurations, and this problem can become potentially severe in the proximity of critical points; in addition, every new set of couplings requires a new chain to be thermalized again. By contrast, in stochastic evolutions, each measurement is \emph{independent} from the others, but involves several Monte~Carlo updates that ``interpolate'' from the prior distribution to the distribution defined by the target coupling(s). The computational cost of this might seem very high at first, due to the relatively large number of Monte~Carlo updates needed for convergence. However, it is crucial to note that, as a byproduct of this procedure, all intermediate couplings can be sampled at the same time. It is then clear that this method is particularly suitable when a fine scan of the parameter space of a theory is required. A typical example is the study of thermodynamics in strongly interacting quantum field theories. Indeed, a full-scale numerical study using this approach has already been reported: it is the high-precision computation of the equation of state in the $\SU(3)$ Yang-Mills theory in $3+1$ dimensions~\cite{Caselle:2018kap}. In that case, the only parameter changed by the protocol is the inverse coupling $\beta$, which in turn also controls the temperature of the system. The major difference with respect to the stochastic evolutions used in this work is that the generator of starting configurations was not a treatable (e.g. normal) prior distribution, but a distribution obtained from a Markov chain thermalized at a certain value $\beta_0$ of the inverse coupling.

We also showed that a common framework can describe in a natural way both stochastic and normalizing flows, following the work of ref.~\cite{Wu:2020snf} and expanding it so to explicitly include stochastic transformations based on Jarzynski's equality. The construction of stochastic normalizing flows is a direct consequence of this connection: a ``hybrid'' flow combining deterministic and stochastic transformations proved to be highly efficient at sampling the target distribution, with relatively simple and short training procedures and a limited number of Monte~Carlo updates. Our previous work on stochastic evolutions suggested that increasing the number of intermediate steps, while expensive per se, is an efficient way of improving the effectiveness of the flow, and this is confirmed by the numerical tests performed in this work. However, it was not clear whether this strategy would work at all for stochastic normalizing flows. Interestingly, this proved to be true also in the latter case and with surprisingly high efficiency. We stress though, that this was observed in a rather specific setup in which a) affine blocks based on CNNs are used, b) stochastic and affine blocks are placed in an alternating order, and c) the number of stochastic and affine blocks is increased roughly in the same manner.

The strong similarities between purely stochastic evolutions and SNFs suggest that the success of the former in full-scale numerical calculations can be replicated with the latter using similar strategies. Moreover, due to the existence of an explicit protocol determined by the stochastic updates, the role of intermediate affine layers can be interpreted straightforwardly. As pointed out above, each stochastic layer can be used to sample the intermediate parameters defined by the protocol $\eta(t)$: the (deterministic) affine layers inserted between them are then trained to ``glue'' together the various steps of the protocol without resorting to further (generally more expensive) intermediate Monte~Carlo updates. 

Before concluding, we would like to point out that the training of SNFs is not necessarily more efficient in general. A standard normalizing flow is intuitively more ``free'' to seek the best possible path between the prior and the target distributions. On the other hand, stochastic normalizing flows are ``constrained'' by the protocol chosen for Monte~Carlo updates; these intermediate steps happen at fixed values in the parameter space of the theory, which the training is ``forced'' to go through. Naively, one could expect that a standard normalizing flow will \emph{eventually} outperform a stochastic one, given the same neural-network architecture; however, this might not happen in a reasonable training time, and a fixed protocol might lead in some instances to a faster training. More work is needed, for example, to understand how the training times needed by SNFs to reach a plateau in the loss function behave when changing the volume of the system.

Among possible directions of future work, our primary interest is to study the effectiveness of SNFs in systems close to criticality, in order to develop the most suitable strategy for SNFs in this region of the parameter space of a theory. More generally, an analysis of the interplay between Monte~Carlo updates and different types of neural-network architectures would be highly insightful and could help one understand what exactly the neural networks are learning when ``coupled'' to Monte~Carlo algorithms in this way. Natural extensions of this work include the use of convolutional architectures for gauge equivariant flows~\cite{Kanwar:2020xzo, Boyda:2020hsi}, rational quadratic splines~\cite{DelDebbio:2021qwf} and continuous equivariant flows~\cite{deHaan:2021erb}.

\vskip1.0cm 
\noindent{\bf Acknowledgements}\\

We thank Kim~Nicoli and Paolo~Stornati for helpful discussions. The numerical simulations were run on machines of the Consorzio Interuniversitario per il Calcolo Automatico dell'Italia Nord Orientale (CINECA). We acknowledge support from the SFT Scientific Initiative of INFN. This work was partially supported by the “Departments of Excellence 2018–2022” Grant awarded by the Italian Ministry of Education, University and Research (MIUR) (L.232/2016). Part of the numerical functions used in the present work are based on ref.~\cite{Albergo:2021vyo}.

\appendix

\section{Derivation of Jarzynski's equality for Monte~Carlo algorithms}
\label{sec:jarzynski_derivation}

\setcounter{equation}{0}

We first set our notation. For a system in thermodynamic equilibrium at temperature $T$, the statistical distribution of the $\phi$ configurations is the Boltzmann distribution $\pi$:
\begin{equation}
\label{Boltzmann_distribution}
\pi[\phi] = \frac{1}{Z} \exp \left( - H[\phi]/T \right).
\end{equation}
The partition function $Z$ is related to the free energy $F$ via
\begin{equation}
\label{partition_function_and_free_energy}
Z = \exp \left( - F/T \right).
\end{equation}
Let $P[\phi\to\phi^\prime]$ denote the normalized conditional probability of a transition from a configuration $\phi$ to a configuration $\phi^\prime$ which defines the Markov-chain algorithm of the Monte~Carlo simulation under consideration. At equilibrium, the Boltzmann distribution has to be stationary: the probability that the system evolves from some configuration $\phi$ to a given configuration $\phi^\prime$ must be equal to the probability that it evolves from $\phi^\prime$ to some other configuration, i.e.
\begin{equation}
\label{Boltzmann_distribution_stationarity}
\sum_\phi \pi[\phi] P[\phi\to\phi^\prime] = \sum_\phi \pi[\phi^\prime] P[\phi^\prime \to \phi].
\end{equation}
A sufficient (albeit not necessary) condition to enforce the validity of eq.~(\ref{Boltzmann_distribution_stationarity}) is to assume that the summands, not only the sums, are equal:
\begin{equation}
\label{detailed_balance}
\pi[\phi] P[\phi\to\phi^\prime] = \pi[\phi^\prime] P[\phi^\prime \to \phi],
\end{equation}
i.e. the detailed-balance condition.

Let us consider an out-of-equilibrium evolution of the system during the time interval from $\tin$ to $\tfin$, denoting the $\tfin-\tin$ difference as $\Delta t$, and assuming that this time interval is divided into $N$ subintervals (which we take to be of equal width $\tau=\Delta t /N$, for the sake of simplicity), setting $t_n=\tin + n \tau$ for integer $0 \le n \le N$. We identify the discrete time steps $t_n$ with the steps in Monte~Carlo time in a Markov-chain algorithm. 

Finally, let us introduce the quantity $\mathcal{R}_N[\phi]$ defined as
\begin{equation}
\label{discretized_exponential_work}
\mathcal{R}_N[\phi] = \exp \left( - \frac{1}{T}\sum_{n=0}^{N-1} \left\{ H_{\eta_{n+1}}\left[\phi_n\right] - H_{\eta_n}\left[\phi_n\right] \right\}\right),
\end{equation}
which represents the sum of the exponentiated work (divided by $T$) done on the system during each of the time intervals of width $\tau$, when the couplings are switched from $\eta_n$ to $\eta_{n+1}$. In the $N \to \infty$ limit, $\mathcal{R}_N[\phi]$ tends to the quantity that is averaged over on the right-hand side of eq.~(\ref{Jarzynski_equality}). Using eq.~(\ref{Boltzmann_distribution}), $\mathcal{R}_N[\phi]$ can be rewritten in terms of the Boltzmann distribution as
\begin{equation}
\label{discretized_exponential_work_Z_pi_ratios}
\mathcal{R}_N[\phi] = \prod_{n=0}^{N-1} \frac{Z_{\eta_{n+1}} \pi_{\eta_{n+1}}\left[\phi_n \right]}{Z_{\eta_n} \pi_{\eta_n}\left[\phi_n \right]},
\end{equation}
so that the average of eq.~(\ref{discretized_exponential_work_Z_pi_ratios}) over all possible trajectories from $\tin$ to $\tfin$ can be written as
\begin{equation}
\label{averaged_discretized_exponential_work_Z_pi_ratios}
\overline{\exp \left( -W/T \right)} = \lim_{N \to \infty} \sum_{\left\{ \phi_n \right\}_{n=0}^N }\pi_{\etain}\left[ \phi_0 \right] \prod_{n=0}^{N-1} \left\{ \frac{Z_{\eta_{n+1}}}{Z_{\eta_n}} \cdot \frac{\pi_{\eta_{n+1}}\left[\phi_n\right]}{\pi_{\eta_n}\left[\phi_n\right]} \cdot P_{\eta_{n+1}}\left[ \phi_n \to \phi_{n+1} \right] \right\},
\end{equation}
having used the fact that the system is initially in thermal equilibrium, hence the probability distribution for the configurations at $t=\tin$ is given by eq.~(\ref{Boltzmann_distribution}), and having denoted the $N+1$ sums over configurations at $\tin$, $t_1$, $t_2$, $\dots$, $t_{N-1}$, $\tfin$ as
\begin{equation}
\sum_{\left\{ \phi_n \right\}_{n=0}^N } = \sum_{\phi_0} \sum_{\phi_1}  \sum_{\phi_2} \dots \sum_{\phi_{N-1}} \sum_{\phi_N}.
\end{equation}
The product of ratios of partition functions in eq.~(\ref{averaged_discretized_exponential_work_Z_pi_ratios}) simplifies to $Z_{\etafin}/Z_{\etain}$. Moreover, using eq.~(\ref{detailed_balance}), the sum appearing on the right-hand side of eq.~(\ref{averaged_discretized_exponential_work_Z_pi_ratios}) can be rewritten as
\begin{equation}
\label{simplified_discretized_exponential_work_pi_ratios}
\overline{\exp \left( -W/T \right)} = \frac{Z_{\etafin}}{Z_{\etain}} \lim_{N \to \infty} \sum_{\left\{ \phi_n \right\}_{n=0}^N }\pi_{\etain}\left[ \phi_0 \right] \prod_{n=0}^{N-1} \left\{\frac{\pi_{\eta_{n+1}}\left[\phi_{n+1}\right]}{\pi_{\eta_n}\left[\phi_n\right]} \cdot P_{\eta_{n+1}}\left[ \phi_{n+1} \to \phi_n \right] \right\}.
\end{equation}
In the latter expression, the ratios of Boltzmann distributions simplify to $\pi_{\etafin}\left[ \phi_N \right]/\pi_{\etain}\left[ \phi_0 \right]$, which, in turn, simplifies against the $\pi_{\etain}\left[ \phi_0 \right]$ factor:
\begin{equation}
\label{discretized_exponential_P_product}
\overline{\exp \left( -W/T \right)} = \frac{Z_{\etafin}}{Z_{\etain}}  \lim_{N \to \infty} \sum_{\left\{ \phi_n \right\}_{n=0}^N }\pi_{\etafin}\left[ \phi_N \right] \prod_{n=0}^{N-1} P_{\eta_{n+1}}\left[ \phi_{n+1} \to \phi_n \right].
\end{equation}
The sum over the initial configurations can be performed explicitly, as $\phi_0$ appears only in the $P_{\eta_1}\left[ \phi_1 \to \phi_0\right]$ term, and the result is $1$, due to the normalization of the conditional transition probability. Next, the same argument can be repeated to sum over the $\phi_1$, $\phi_2$, $\dots$, $\phi_{N-1}$ configurations. Finally, noting that also $\pi_{\etafin}\left[ \phi_N \right]$ is normalized to $1$, one obtains
\begin{equation}
\label{final_discretized_exponential_P_product}
\overline{\exp \left( -W/T \right)} = \frac{Z_{\etafin}}{Z_{\etain}},
\end{equation}
which is eq.~(\ref{Jarzynski_equality}).

We remark that, although in this proof we used the Boltzmann distributions at all times $\tin \le t \le \tfin$, we did this only to re-express the $\exp(-H/T)$ terms appearing in eq.~(\ref{discretized_exponential_work}). The configurations $\phi_n$ at $t > \tin$ are \emph{not} in thermal equilibrium. Moreover, for simplicity, we assumed the temperature $T$ to be constant throughout the evolution of the system along each trajectory, but this does not necessarily have to be the case~\cite{Chatelain:2007ts}.

\bibliography{paper}

\end{document}